\documentclass[12pt,preprint]{aastex}
\shorttitle{\ion{O}{6} Absorption in AGN Host Galaxy}
\shortauthors{Bonamente \& Dixon}

\newcommand{\eg}{e.g.}

\newcommand{\fig}[1]{Fig.~\ref{#1}}

\newcommand{\hone}{\ion{H}{1}}
\newcommand{\htwo}{H$_2$}

\newcommand{\osix}{\ion{O}{6}}

\newcommand{\kms}{km s$^{-1}$}

\newcommand{\specfit}{{\small SPECFIT}}

\newcommand{\euve}{{\it EUVE}}
\newcommand{\fuse}{{\it FUSE}}

\newcommand{\rosat}{{\it ROSAT}}
\newcommand{\heao}{{\it HEAO-1}}

\newcommand{\source}{2MASX~J21362313-6224008}

\begin{document}
 
\title{FUSE Observations of Galactic and Intrinsic Absorption
in the Spectrum of the Seyfert 1 Galaxy 2MASX~J21362313-6224008\/\footnotemark}

\footnotetext[1]{Based on observations made with the NASA-CNES-CSA
{\it Far Ultraviolet Spectroscopic Explorer.  FUSE}\/ is operated for NASA by
the Johns Hopkins University under NASA contract NAS5-32985.}

\author{Massimiliano Bonamente}
\affil{Physics Department, The University of Alabama in Huntsville,
John Wright Dr., OB201, Huntsville, AL 35899}
\email{bonamem@email.uah.edu}

\and

\author{W.\ Van Dyke Dixon}
\affil{Department of Physics and Astronomy, The Johns Hopkins University,
3400 North Charles Street, Baltimore, MD 21218} 
\email{wvd@pha.jhu.edu}

\begin{abstract}

We present the far-ultraviolet spectrum of the Seyfert 1 galaxy \source\/ obtained
with the {\it Far Ultraviolet Spectroscopic Explorer (FUSE)}.
The spectrum features absorption from Galactic \ion{O}{6} at two velocities and 
redshifted \hone\ Lyman $\beta$ and $\gamma$, \ion{C}{2}, \ion{C}{3}, and
\ion{O}{6}.
The redshifted absorption features represent a single kinematic component 
blueshifted by $\sim 310$ km s$^{-1}$ relative to the AGN. We 
use photoionization models to derive constraints on the physical parameters of the absorbing gas.
An alternative interpretation for the absorption lines is also proposed, whereby the
absorbing gas is associated with an intervening galaxy cluster.

\end{abstract}


\section{INTRODUCTION}
A large fraction of Seyfert 1 galaxies exhibits intrinsic UV absorption associated
with the active galactic nucleus (AGN).  The absorption lines are commonly
blueshifted with respect 
to the source, indicating that the absorbing material undergoes a net radial 
outflow (\eg, \citealt{Crenshaw:99}).  The absorbing material is interpreted
as photoionized gas with ionization parameter $U \sim 0.01-0.1$ and 
total hydrogen column density $N_H \sim 10^{18}-10^{21}$ cm$^{-2}$ 
\citep{Kriss:00,Kriss:03,Romano:02}.

\begin{figure}
\plotone{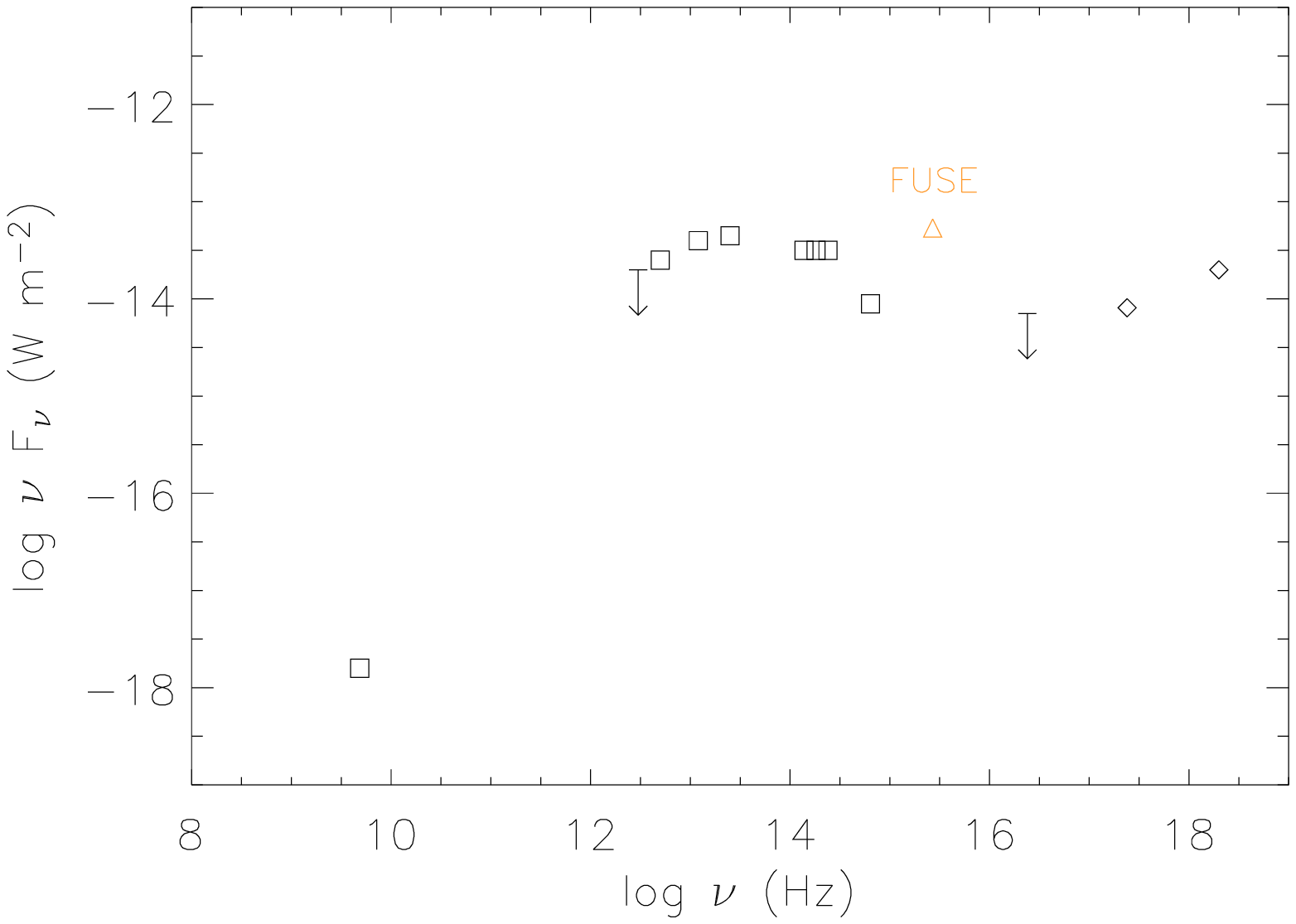}
\caption{Spectral energy distribution of  \source.
To the right of the \fuse\/ data point are the \euve\/, \rosat\/ and \heao\/ measurements, respectively.  The lower-frequency measurements were derived from the available
literature through the NASA/IPAC Extragalactic Database
(http://nedwww.ipac.caltech.edu).
\label{sed}}
\end{figure}

In this paper we present the results of a 37 ks observation of \source, a
Seyfert 1 galaxy with a measured redshift $z=0.0588$ \citep{Hewitt:91}
and visual magnitude V=15.2 mag \citep{Remillard:86}, with the
{\it Far Ultraviolet Spectroscopic Explorer (FUSE)}.
The available X-ray HEAO-1 data  
of the source (also identified as 1H2129-624 
and H2132-626) were analyzed by \citet{Remillard:86}, who derive a luminosity
$L_{2-10\; {\rm keV}}=3 \times 10^{44}$ erg s$^{-1}$,
which is on the bright end of the quasar luminosity function \citep{George:00}.
The object was observed in soft X-rays by \rosat; the \rosat\/ All-Sky Bright Source Catalogue \citep{Voges:99} and
the \rosat\/ Bright Survey \citep{Schwope:00} detect a strong X-ray source at the
location of our target, which is identified as 1RXS J213623.1-622400.
(Its quoted X-ray flux may be contaminated by up to 10\% by the faint source
1RXS J213530.1-623005 approximately 12 arcmin to the southwest.)
\euve\/ observed the source during its  All-Sky Survey, but the short exposure time
resulted only in an upper limit \citep{Marshall:95}.
The spectral energy distribution (SED) of \source\ is shown in Fig. \ref{sed}.

\section{OBSERVATIONS AND DATA REDUCTION}

\fuse\/ consists of four separate optical systems.  Two employ LiF
coatings and are sensitive to wavelengths from 990 to 1187 \AA, while
the other two use SiC coatings, which provide reflectivity to
wavelengths as short as 905 \AA.  The four channels overlap between 990
and 1070 \AA.  For a complete description of the instrument, see \citet{Moos:00}
and \citet{Sahnow:00}.

\footnotetext[2]{{\it The CalFUSE Pipeline Reference Guide}\/ is available at
\url{http://fuse.pha.jhu.edu/analysis/pipeline\_reference.html}.}

The \fuse\/ spectrum of 2MASX~J21362313-6224008 (data set D9030401) was
obtained in 10 separate exposures on 2003 May 25-26.  The total
integration time was 37 ks, of which 12 ks were obtained during orbital
night.  All observations were made through the 30\arcsec\ $\times$
30\arcsec\ (LWRS) aperture.  The data were reduced using version 2.4 of
the CalFUSE calibration software pipeline, described in {\it The
CalFUSE Pipeline Reference Guide}\/\footnotemark\
\citep{calfuse_pipeline_reference}, but with the following
modification:  The first half of the pipeline, which corrects for
time-dependent effects (such as spacecraft jitter) was run separately
on each exposure.  The resulting position-corrected photon-event lists
were combined, using the program {\small TTAG\_COMBINE}, into a single
data file.  The second half of the pipeline, which performs background
subtraction and spectral extraction (among other tasks), was run on the
combined data file.  By thus using the entire data set to scale the
background model, we optimize its fidelity, an important consideration
for faint continuum sources.

\begin{figure}
\plotone{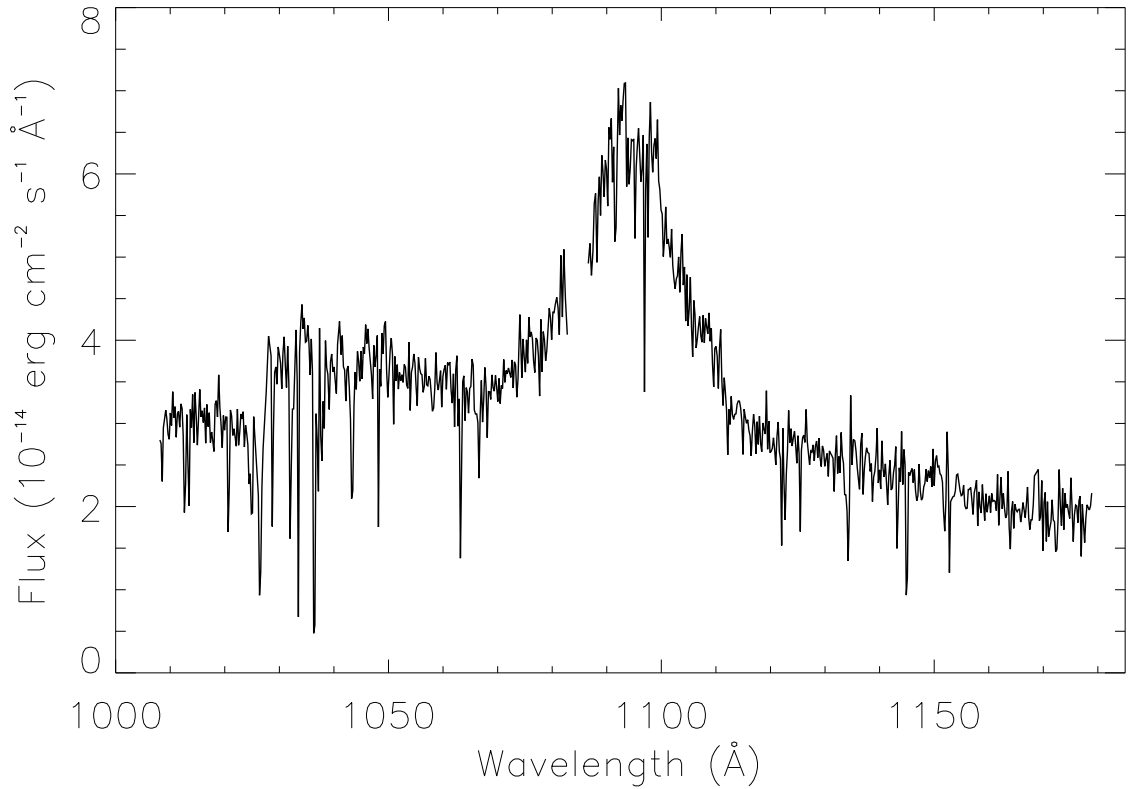}
\caption{\fuse\ spectrum of the Seyfert 1 galaxy \source.
The data are smoothed by 32 pixels, or about four resolution elements, and
geocoronal emission lines (defined as resolved features extending more than three
standard deviations above the local median) are excluded.
Wavelengths around 1085 \AA\ are not covered by the LiF channels.
The spectrum exhibits a power-law continuum with a spectral break at the
wavelength of Galactic Lyman $\beta$.  At shorter wavelengths, the
spectrum is flat to the Lyman limit.  The broad emission feature is
redshifted \osix\ emission.  Most absorption lines are due to the ISM of
our Galaxy.
\label{full_spectrum}}
\end{figure}

The \fuse\/ flux calibration, based on theoretical models of
white-dwarf stellar atmospheres, is believed accurate to about 10\%
\citep{Sahnow:00}.  Error bars are assigned to the data assuming
Gaussian statistics.  To increase their signal-to-noise ratio, we bin
the spectra by four detector pixels, about half of a resolution
element.  The FUV spectrum of \source\ (\fig{full_spectrum}) 
shows a power-law continuum with a broad \osix\ emission feature 
at the redshift of the AGN.
Selected spectral regions showing redshifted absorption
features are presented in \fig{absorption}.

\section{SPECTRAL ANALYSIS \label{spectral}}

\begin{figure}
\epsscale{0.75}
\plotone{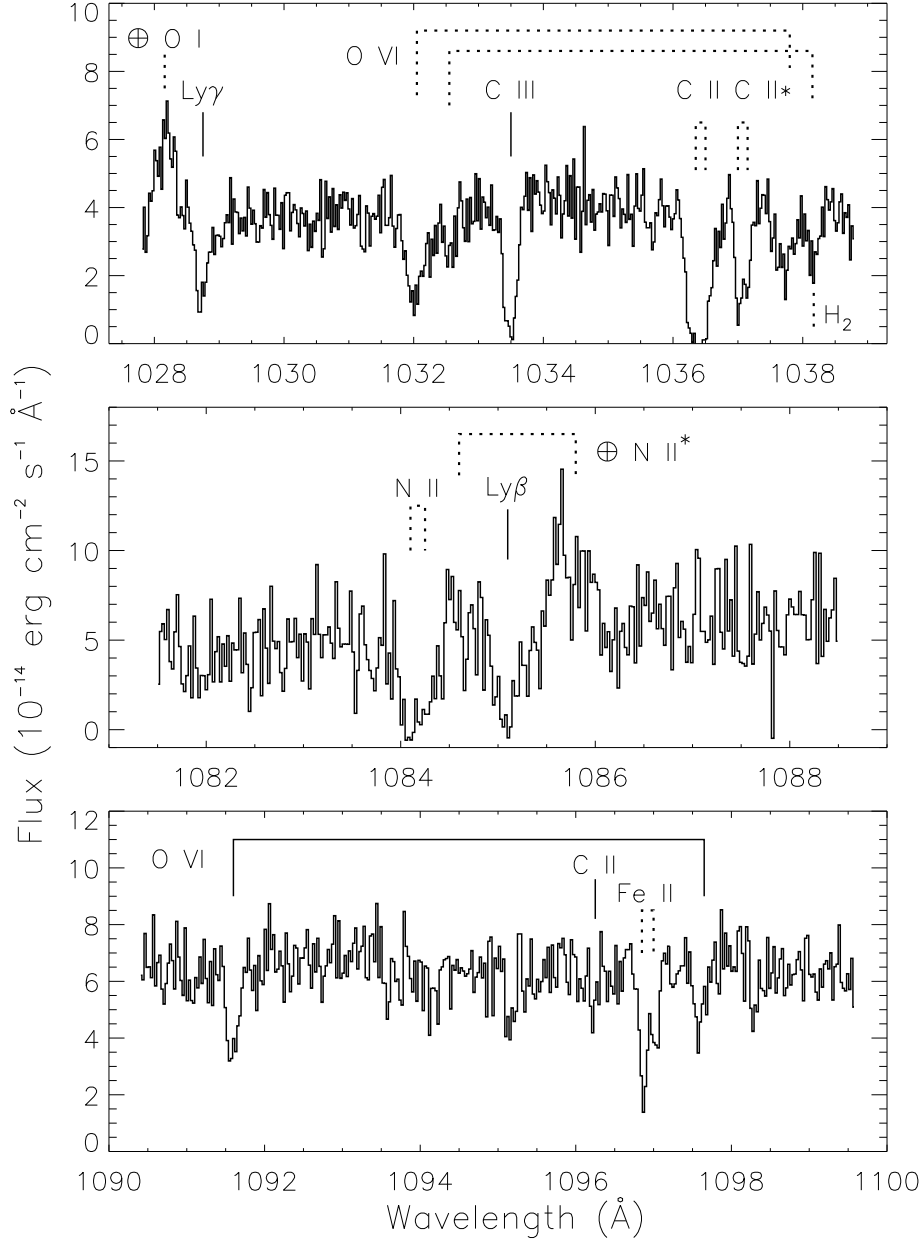}
\caption{
Redshifted absorption features in the \fuse\ spectrum of 2MASX~J21362313-6224008.
Interstellar absorption and geocoronal emission features are marked with dotted lines,
redshifted absorption features with solid lines.
{\it Top:}\/ The Ly$\gamma$ $\lambda 973$ and \ion{C}{3} $\lambda 977$
features are blueshifted by 310 \kms\ relative to the AGN.  The Galactic \ion{O}{6}
$\lambda \lambda 1032, 1038$, \ion{C}{2} $\lambda 1036$, and \ion{C}{2}*
$\lambda 1037$ features show two velocity components.
{\it Middle:}\/ Redshifted Ly$\beta$ absorption falls
between the geocoronal \ion{N}{2} $\lambda \lambda 1094.6, 1085.7$ emission
lines.  {\it Bottom:}\/ Redshifted \ion{O}{6} $\lambda \lambda 1032,
1038$ and \ion{C}{2} $\lambda 1036$ with Galactic \ion{Fe}{2} $\lambda
1097$ (two velocity components).  The apparent absorption features at
1095.2 and 1098.5 \AA\ are detector artifacts.
\label{absorption}}
\end{figure}

Because the four channels are essentially independent spectrographs,
they have different line-spread functions, and their data cannot be
safely combined into a single spectrum.  Instead, we use only the
spectrum from the channel with the highest sensitivity at the
wavelength of interest.  We identify statistically-significant
absorption features with a simple routine that bins each spectrum to
the instrument resolution and flags regions whose flux lies more than
three standard deviations below the local median.  Most of these
features are due to the interstellar medium (ISM) of our Galaxy, but a
handful share the redshift of the AGN:
Ly$\gamma$ $\lambda 973$, \ion{C}{3} $\lambda 977$, Ly$\beta$ $\lambda
1026$, \ion{O}{6} $\lambda \lambda 1032, 1038$, and \ion{C}{2} $\lambda
1036$ (\fig{absorption}).

Absorption-line profiles are modeled with the interstellar
line-fitting package written by M. Hurwitz and V. Saba.  Wavelengths,
oscillator strengths, and other atomic data are taken from
\citet{Morton:91}.  Given a column density and Doppler broadening
parameter, the program computes a Voigt profile for each absorption
feature and produces a high-resolution (0.001 \AA) spectrum of $\tau$
versus wavelength.  The model spectra are convolved with a Gaussian of
FHWM = 0.08 \AA, roughly the spectral resolution of our data,
and rebinned to 0.01 \AA.

\footnotetext[3]{The Image Reduction and Analysis Facility (IRAF) is
distributed by the National Optical Astronomy Observatories, which is
operated by the Association of Universities for Research in Astronomy,
Inc., (AURA) under cooperative agreement with the National Science
Foundation.}

Model spectra are fit to the data using the nonlinear curve-fitting
program \specfit\ \citep{Kriss:94}, which runs in the
IRAF\footnotemark\ environment.  \specfit\ performs a $\chi^2$
minimization of the model parameters.  Error bars for a particular
parameter are derived by fixing that parameter at the best-fit value,
then raising it, allowing the other model parameters to vary freely,
until $\chi^2$ is increased by 1, which corresponds to a 1 $\sigma$
deviation for a single interesting parameter \citep{Avni:76}.

We begin with the segment of the LiF~1A spectrum shown in the top panel
of \fig{absorption}, which includes the redshifted Ly$\gamma$ $\lambda
973$ and \ion{C}{3} $\lambda 977$ lines and Galactic \osix\ $\lambda
\lambda 1032, 1038$ absorption at two velocities.  Comparison with data
taken during orbital night (not shown) indicates that Ly$\gamma$ is
well separated from the geocoronal \ion{O}{1} $\lambda 1028$ emission
line.  The redder \osix\ $\lambda 1038$ line is contaminated by $J$=1
\htwo\ absorption.  Assuming a linear continuum over the wavelength
region 1028.4--1038.6 \AA, we model all nine species/velocity
components labeled in \fig{absorption}.  We constrain the depth of the
\htwo\ line by fitting two other $J$=1 \htwo\ lines in the 1049--1052
\AA\ range simultaneously.  Derived parameters for the Galactic
\osix\ lines are presented in Table \ref{osix_lines} and those for the
redshifted features in Table \ref{red_lines}.

To model the redshifted Ly$\beta$ $\lambda 1026$ feature, we use data
from the SiC~1A channel (middle panel of \fig{absorption}), as the more
sensitive LiF channels do not include this wavelength range.  The
Ly$\beta$ line falls between a pair of geocoronal emission features due
to \ion{N}{2}* $\lambda \lambda 1084.6, 1087.5$.  The pair of
\ion{N}{2} $\lambda 1084$ absorption features is interstellar. A
comparison with the night-only spectrum (not shown) indicates that
\ion{N}{2}* $\lambda 1084.6$ does not contribute significantly to
this spectrum and that the narrow peak on the blue shoulder of the
\ion{N}{2}* $\lambda 1087.5$ line is probably not geocoronal, but
intrinsic to the target spectrum.  Diffuse emission filling the LWRS
aperture yields a line profile that is well approximated by a top-hat
function with a width of $\sim 106$ \kms.  The \ion{N}{2}* $\lambda 1087.5$ 
line is even broader, which also suggests that a second emission component is
present.  We model the region between 1081 and 1088 \AA\ with a linear
continuum, a pair of interstellar \ion{N}{2} features, and a redshifted
Ly$\beta$ line.  We model the emission feature with a 106
\kms\ top hat and a narrow Gaussian.  Derived parameters for the
redshifted Ly$\beta$ line are presented in Table \ref{red_lines}.

Finally, we turn to the redshifted \ion{O}{6} $\lambda \lambda 1032,
1038$ and \ion{C}{2} $\lambda 1036$ features in the LiF~2A spectrum
(bottom panel of \fig{absorption}).  These features fall on the broad
peak of the redshifted \osix\ emission feature, which is well fit by a linear
continuum.  The apparent absorption features at 1095.2 and 1098.5 \AA\ 
are detector artifacts.  The pair of \ion{Fe}{2} $\lambda
1097$ lines represents absorption from two velocity components in the
ISM of our Galaxy.  This velocity structure is repeated in \ion{Fe}{2}
lines at longer wavelengths.  As with the Galactic \osix\ lines, we fit
both components of the \ion{O}{6} doublet simultaneously and quote a
single set of derived parameters in Table \ref{red_lines}.

If the absorbing cloud(s) cover only part of the emitting region, then our
derived column densities are lower limits.  Alternatively, if the absorption features
that we have fit with a single velocity component are due to multiple unresolved
clouds, then our derived column densities are upper limits.  To investigate
the importance of these effects, we calculate the covering fraction of the \osix 
-absorbing gas as a function of velocity $C_f (v)$ by comparing the depths
of the two components with their expected 2:1 ratio.  Following the recipe given by
\citet{Hamann:97b}, we find that $C_f$ for the \osix -absorbing gas ranges from
0.5 to 1.0 across the line.  Though the error bars are large, this range of values
suggests that multiple velocity components are partially resolved in our data.
A second velocity component fit to the \osix\ doublet is, however, significant
at only the 2 $\sigma$ level.  We therefore assume a single velocity component
with a covering fraction $0.5 \leq C_f \leq 1.0$ in the analysis that follows.

Misalignments among the four \fuse\ channels can lead to offsets in the
wavelength scales of their spectra.  To correct for such offsets, we
measure the positions of ISM features in each spectrum and adjust their
velocities to a common scale.  Our standard is the LiF~1A spectrum, as
data from this channel are used to guide the spacecraft.  Specifically,
we measure the \ion{Ar}{1} $\lambda \lambda 1048, 1067$ lines in the
LiF~1A spectrum, \ion{Ar}{1} $\lambda 1048$ in SiC~1A, and \ion{Fe}{2}
$\lambda 1097$ in LiF~2A.  Absolute velocities remain uncertain,
however, because an offset in the position of the target relative to
the center of the LWRS aperture can lead to a zero-point uncertainty in
the \fuse\ wavelength scale of up to $\pm 0.15$ \AA.  This uncertainty
is not included in our quoted error bars.
                                                                                                    
Though a detailed analysis of intrinsic emission in the spectrum of
\source\ is beyond the scope of this paper, we can provide a brief
description of the salient features.  
At long wavelengths, the continuum follows a power law
of the form $f \propto \lambda^{-\alpha}$, where $\alpha = 5.3 \pm 0.2$.
There is a spectral break at the wavelength of Galactic
Lyman $\beta$, and the spectrum continues flat to the Lyman limit.
The \osix\ feature that dominates \fig{full_spectrum} is well fit by a
pair of broad \osix\ emission lines (FWHM = $6280 \pm 160$ \kms) and a
pair of narrow \osix\ emission lines (FWHM = $1060 \pm 150$ \kms),
where all lines are fit with Gaussian profiles.  Our target is similar
to NGC~3783 in that broad emission from Ly $\beta$ and higher-order
Lyman lines is negligible \citep{Gabel:03}.

\section{DISCUSSION}

Derived parameters for the Galactic \osix\ lines are presented in Table
\ref{osix_lines}, and those for the redshifted features in Table
\ref{red_lines}.  
The Lyman $\beta$ and
$\gamma$ lines  yield consistent values for the redshifted \hone\ column
density and Doppler parameter $b$.  We adopt a value of
$N($\hone$) = 10^{15}$ cm$^{-2}$ in our analysis.

\subsection{Photoionization modeling \label{phot_model}}

The redshifted \ion{H}{1}, \ion{C}{2}, \ion{C}{3}, and \ion{O}{6} lines
indicate the presence of ionized gas in the neighborhood
of the AGN (\eg, \citealt{Crenshaw:99}) moving away from the central
source with a speed of $\sim$ 310 km s$^{-1}$ (Table \ref{red_lines}).
Photoionization models can be used to constrain the parameters of the
absorbing gas. We use Cloudy v.94 \citep{Ferland:96} to calculate
the fractional abundance $f_{ion}$ of an element in a given ionization
state, assuming the illuminating continuum described by the Cloudy ``table agn''
model, an absorbing gas with solar abundances and total hydrogen density
$n($H$)=10^5$ cm$^{-3}$  and a grid of values for the
total hydrogen column density $N_H$\footnote{The hydrogen density $n($H)
and column density $N_H$
include atomic and molecular hydrogen in all ionization stages, as defined
in the Cloudy v.94 manual \citep{Ferland:96}. In the present model, hydrogen is 
mostly in atomic form.}
and ionization parameter $U$.
The measured SED of \source\/ (\fig{sed}) is consistent with
the ``table agn'' model employed in the calculation; other studies (\eg, \citealt{Romano:02})
indicate that the use of other SEDs for the illuminating spectrum results in only
minor changes to the derived values of $N_H$ and $U$.
We compare the column densities $N_{ion}$ presented in
Table \ref{red_lines} with the predictions of the
photoionization model to constrain $U$ and $N_H$ in the absorbing gas.

\begin{figure}
\rotatebox{90}{
\plotone{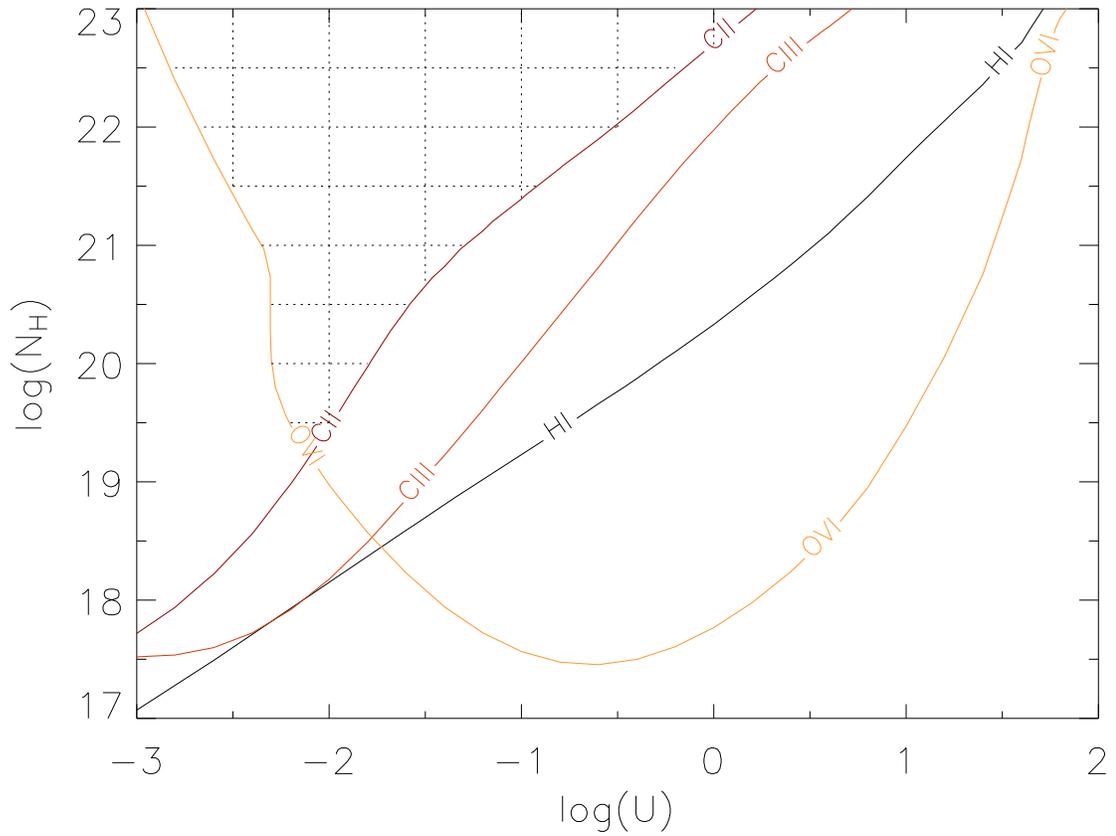}}
\caption{Photoionization curves at constant ionic column density. $U$ is the ionization parameter and $N_H$ the total hydrogen column density. The column densities are from Table \ref{red_lines}; the \ion{H}{1} curve represents a neutral hydrogen 
column density of $10^{15}$ cm$^{-2}$. The cross-hatched area is the region where all constraints are met.
\label{photo}}
\end{figure}

In Section 3, the covering fraction of the absorbing gas in \source\ was found
to be $0.5 \leq C_f \leq 1.0$, consistent with
most other intrinsic UV absorbers in AGN \citep{Crenshaw:99}.
Following \citet{Arav:01} and \citet{Romano:02}, we plot
curves of constant $N_{ion}$ in the log($U$)--log($N_H$) plane
in Fig. \ref{photo}.  If the absorbing gas covers only part of the emitting region, then
$N_{ion}$ is a lower limit and the allowed region of parameter space lies {\it above} 
all the curves in Fig. \ref{photo} (see \citealt{Arav:01} for details). The total
hydrogen column density of warm absorbers in Seyfert galaxies is usually
lower than $ 10^{21}$ cm$^{-2}$ \citep{Kriss:00,Kriss:03,Romano:02,Blustin:03}.
Given the constraints of Fig. \ref{photo}, we tentatively place the 
UV absorber at $\log (U) = -2$, $\log (N_H) = 20$, keeping in mind
that substantially higher column densities are not excluded by our data.
Observations of other UV, optical, or X-ray lines are required
to further constrain the state of the absorber in this AGN.

If the
absorption is due to multiple velocity components, the assumption of a single,
uniform cloud breaks down, and this model is not applicable.



\subsection{Physical condition of the warm absorber}
The ionization parameter is defined as $U=Q(H)/4 \pi r_0^2 n(H) c$, where
$Q(H)$ is the number of ionizing photons ($E \geq$ 1 Ry),
$n(H)$ is the total hydrogen number density, $c$ is the speed
of light, and $r_0$ is the distance of the illuminated face of the cloud from
the central source.
We use the measured 2-10 keV luminosity (3$\times 10^{44}$ erg s$^{-1}$)
in conjunction with the assumed ``table agn'' SED to derive a flux of
$Q(H)=10^{56}$ s$^{-1}$ ionizing photons for \source. 


Knowing $U$, $N_H$, and the density $n(H)$, one can derive the distance,
size, and mass of the absorbing cloud. 
Our data provide no constraints on the density of the absorber --- variability
studies or other information such as imaging data are necessary to measure  $n(H)$
(e.g., \citealt{Hamann:97a,Crenshaw:02}) --- so we will continue to use the
value $n($H$)=10^5$ cm$^{-3}$ assumed by the ``table agn'' model.
Adopting the ``best-guess'' values of $\log (U) = -2$, $\log (N_H) = 20$
derived above,
we find that the illuminated face of the absorber lies  
$r_0 = 160 \; U_{-2}^{-1/2} \; n(H)_{5}^{-1/2}$ pc from the central source.
The size of the absorbing cloud is  simply given by
$L=N_H/n(H)=10 \; N_{{\rm H},20} \; n(H)_{5}^{-1}$ pc.
Since $L << r_0$ for a wide range of hydrogen densities $n(H)$,
the volume of the warm gas can be approximated as $V \sim 4 \pi r_0^2 L \times C_f$,
and the mass of the warm absorber is $M \sim 2.1 \times 10^7  \; U_{-2}^{-1} \;
n(H)_{5}^{-1} \; N_{{\rm H},20} C_f\; M_{\sun}$.

\subsection{An intergalactic origin for the absorption lines?}

The line of sight to \source\ intersects several clusters of galaxies, two of which
have measured redshifts: A3782 at $z=0.0557$ and APMCC684 at $z=0.056$.
The difference in redshift between the measured absorption lines
($z \sim 0.0578$) and the two clusters is so small that an alternative 
interpretation of the absorption lines can be entertained, viz., the source shines
through the foreground cluster(s), which cause the redshifted absorption lines.

The center of A3782 lies $\sim 26$ arcmin from \source, which corresponds to
approximately 1.8 Mpc (for a Hubble constant of $H_0=70$ km s$^{-1}$ Mpc$^{-1}$).
The center of APMCC684 is $\sim$ 17.5 arcmin distant, which corresponds to 1.2 Mpc.
Clusters of galaxies are known to host hot intergalactic gas ($T \geq 10^7$ K) -- too hot to
produce the observed absorption lines -- and oftentimes a lower temperature phase 
($T \sim 10^5-10^6$ K;
\citealt{Lieu:96a,Lieu:96b,Bonamente:02,Bonamente:03})
that could contain substantial amounts of \ion{O}{6}, \ion{C}{2}, \ion{C}{3},
and neutral atomic hydrogen. 
This warm intergalactic gas has been observed in emission in several
clusters (i.e., the ``soft excess'' phenomenon; \citealt{Bonamente:02}; \citealt{Kaastra:03}; \citealt{Nevalainen:03}), 
notably in the Coma cluster,
where it extends some 2.6 Mpc from the cluster center \citep{Bonamente:03}.
Detection of \ion{O}{6} absorption associated with the Local Group of galaxies has been recently
reported by \citet{Nicastro:03}, and detection of an \ion{O}{6} absorption system associated
with another galaxy group was reported by \citet{Tripp:00}. Here we investigate 
the association of the detected absorption systems with the two galaxy clusters A3782 and APMCC684.

\begin{figure}
\plotone{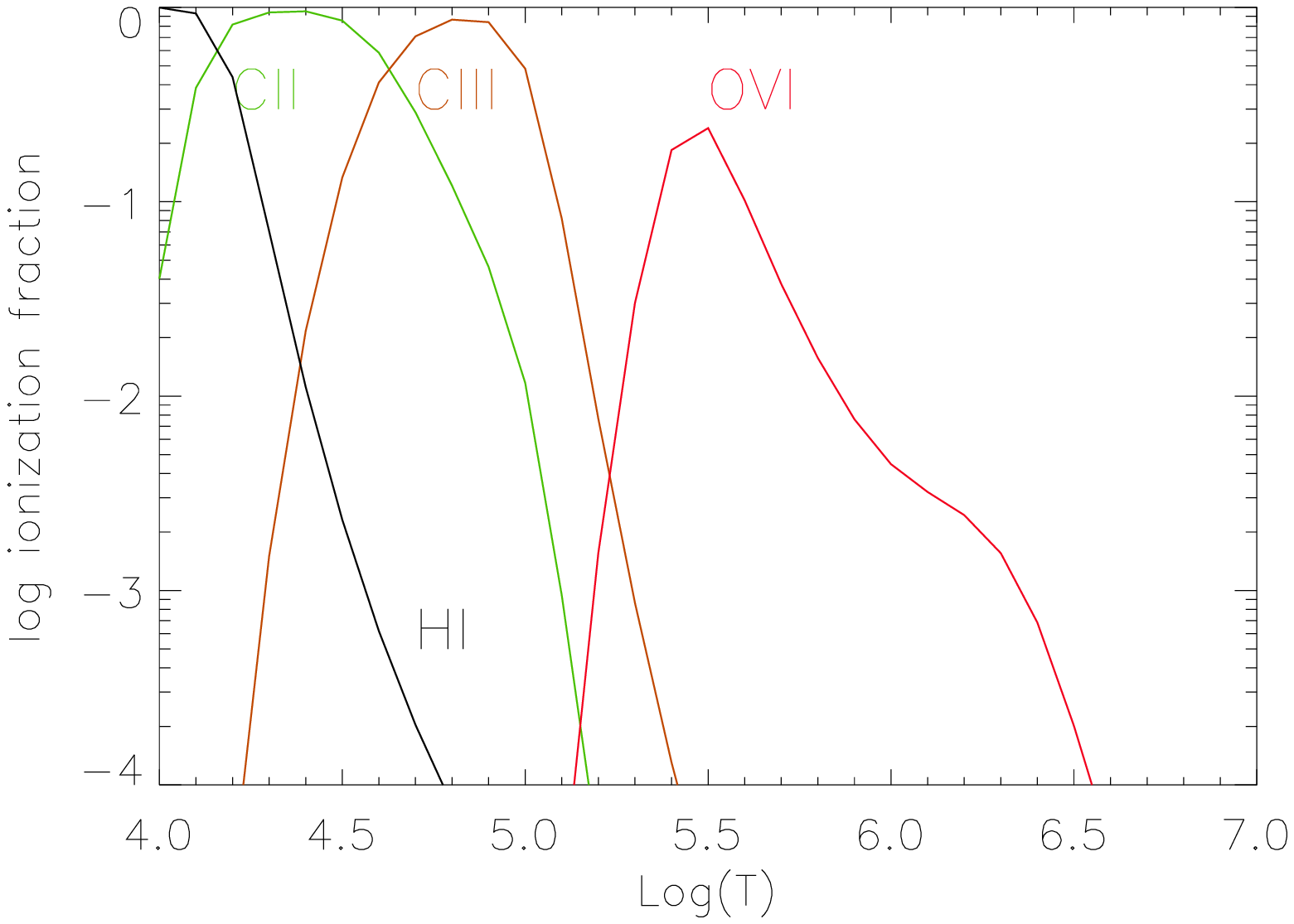}
\caption{Ionization fractions of  \ion{H}{1}, \ion{C}{2}, \ion{C}{3}, and \ion{O}{6} for a gas
in collisional equilibrium (from \citealt{Mazzotta:98}).
\label{mazzotta}}
\end{figure}

We use the soft excess measurements of \citet{Bonamente:03} for the Coma cluster to estimate the
amount of low-ionization gas in clusters of galaxies. Several other
clusters contain warm gas in amounts comparable (within a factor of a few) to those of the Coma cluster.
The warm gas is generally more diffuse than the hot gas \citep{Bonamente:02}, and it
could reside either inside the cluster (i.e., {\it mixed} with the hot gas)
or in filamentary structures outside the cluster, as is often seen in
hydrodynamical simulations (\eg, \citealt{Cen:99}). 
If the warm gas is bound to the cluster, \citet{Bonamente:03} show that it must have a
density of $1.5 \times 10^{-3} - 6 \times 10^{-5}$ cm$^{-3}$ throughout the cluster.
Assuming a cluster radius
of $\sim$3 Mpc, in agreement with current X-ray measurements, this density
implies \ion{H}{1} column densities of $\sim 10^{20}-10^{22}$ cm$^{-2}$.
The metal abundance of the warm gas is
$\sim 0.1$ solar (\citealt{Bonamente:03}); this implies
total carbon column densities of $N_C \sim 5 \times 10^{15}-5 \times 10^{17}$ cm$^{-2}$
and oxygen column densities of $N_O \sim 10^{16}- 10^{18}$ cm$^{-2}$.
Figure \ref{mazzotta} shows the
ionization fractions of the ions of concern as a function of temperature, assuming ionization
equilibrium \citep{Mazzotta:98}. Column densities of 
\ion{H}{1}, \ion{C}{2}, \ion{C}{3}, and \ion{O}{6} predicted by this scenario are
therefore consistent, for a wide range of temperatures,  with those detected toward 
\source~(Table \ref{red_lines}). 
The warm gas may alternatively reside in filamentary structures with densities
of $\sim 10^{-4} - 10^{-5}$ cm$^{-3}$ \citep{Dave:01}.
In this case, \citet{Bonamente:03} show that
the filaments will extend for $\sim$ 10 Mpc outside the cluster, yielding similar
ion column densities.
In addition, the Doppler parameters $b$ of a warm gas at $T \sim 10^4-10^6$ K are 
fully consistent with the values derived in Table 2 \citep{Spitzer:78}, 
and similar to those of \citet{Tripp:00}. 

We conclude that the current data for \source~are consistent with the redshifted
absorption lines of Table \ref{red_lines} originating from warm gas associated with an intervening
galaxy cluster.

\section{CONCLUSIONS}
We detect Galactic \ion{O}{6} absorption in the direction of \source\
and \hone\ Lyman $\beta$ and $\gamma$, \ion{C}{2},
\ion{C}{3}, and \ion{O}{6} absorption at a redshift of $z \sim 0.0578$.
The redshifted absorption lines are consistent with a circumnuclear
absorber outflowing with a relative velocity of $-310$ km s$^{-1}$.
We derive constraints on its physical parameters through photoionization modeling.
Alternatively, we suggest that the redshifted absorption may originate in or around a 
cluster of galaxies  located along the line of sight.

\acknowledgments

This research has made use of NASA's Astrophysics Data System
Bibliographic Services and the NASA/IPAC Extragalactic Database (NED),
which is operated by the Jet Propulsion Laboratory, California
Institute of Technology, under contract with the National Aeronautics
and Space Administration.
The authors thank Pat Romano for insightful discussions and for
sharing the Cloudy commands for the photoionization calculations. 
This work is supported by a \fuse\ Cycle 4 Guest Investigator grant from NASA.

\nocite{Mihalas:Binney:81}


\clearpage 

\bibliographystyle{apj}
\bibliography{apjmnemonic,ms}	


\clearpage

\clearpage 

\begin{deluxetable}{ccc}
\tablewidth{0pt}
\tablecaption{Galactic \ion{O}{6} Absorption Features\label{osix_lines}}
\tablehead{
\colhead{Velocity} & \colhead{Doppler Parameter} & \colhead{Column Density}\\
\colhead{(\kms)} & \colhead{(\kms)} & \colhead{$\log N$ (cm$^{-2}$)}
}
\startdata
$\phn27 \pm 3$ & $53 \pm \phn6$ & $14.44 \pm 0.03$ \\
$175 \pm 6$ & $26 \pm 15$ & $13.77 \pm 0.09$
\enddata

\tablecomments{Velocities are quoted relative to the local standard of
rest (LSR; Mihalas \& Binney 1981).}

\end{deluxetable}




\begin{deluxetable}{llccc}
\tablewidth{0pt}
\tablecaption{Redshifted Absorption Features\label{red_lines}}
\tablehead{
& & & \colhead{Doppler}\\
& & \colhead{Velocity}& \colhead{Parameter} & \colhead{Column Density}\\
\colhead{Feature} & \colhead{Redshift} & \colhead{(\kms)} &
\colhead{(\kms)} &
\colhead{$\log N$ (cm$^{-2}$)}
}
\startdata
Lyman $\gamma$ $\lambda 972.5$ & 0.057760(7) & $-312 \pm 2$ & $27 \pm 5$ & $14.97 \pm 0.05$ \\
\ion{C}{3} $\lambda 977.0$     & 0.057783(5) & $-305 \pm 1$ & $24 \pm 3$ & $13.95 \pm 0.06$ \\
Lyman $\beta$ $\lambda 1025.7$ & 0.057770(15) & $-309 \pm 3$ & $26 \pm 9$ & $14.89 \pm 0.14$ \\
\ion{O}{6} $\lambda \lambda 1031.9,1037.6$ & 0.057783(9) & $-305 \pm 3$ &  $20 \pm 2$ & $13.95 \pm
0.05$ \\
\ion{C}{2} $\lambda 1036.4$ & 0.057760(87) & $-312 \pm 3$  & $6 \pm 10$ & $13.45 \pm 0.16$ \\
\enddata
                                                                                                           
\tablecomments{In column 2, the digits in parentheses represent the one-standard-deviation uncertainty in the final digits of the redshift.  In column 3, velocities are quoted relative to the systemic velocity of \source\ ($z = 0.0588, cz = 17,630$ \kms).}
                                                                                                           
\end{deluxetable}

\end{document}